\documentclass[final]{cvpr}

\usepackage{times}
\usepackage{epsfig}
\usepackage{graphicx}
\usepackage{amsmath}
\usepackage{amssymb}
\usepackage[table,x11names]{xcolor}

\usepackage[pagebackref=true,breaklinks=true,colorlinks,bookmarks=false]{hyperref}

\def\cvprPaperID{****} 
\def\confYear{CVPR 2021}
\definecolor{grayhighlight}{RGB}{213,229,255}

\newsavebox\CBox
\def\textBF#1{\sbox\CBox{#1}\resizebox{\wd\CBox}{\ht\CBox}{\textbf{#1}}}

\begin{document}

\title{Real-Time Quantized Image Super-Resolution on Mobile NPUs,\\ Mobile AI 2021 Challenge: Report}
\author{
Andrey Ignatov \and Radu Timofte \and Maurizio Denna \and Abdel Younes \and Andrew Lek \and
Mustafa Ayazoglu \and
Jie Liu \and Zongcai Du \and
Jiaming Guo \and Xueyi Zhou \and Hao Jia \and Youliang Yan \and
Zexin Zhang \and Yixin Chen \and Yunbo Peng \and Yue Lin \and
Xindong Zhang \and Hui Zeng\and
Kun Zeng \and Peirong Li \and Zhihuang Liu \and Shiqi Xue \and Shengpeng Wang
}

\maketitle

\begin{abstract}

Image super-resolution is one of the most popular computer vision problems with many important applications to mobile devices. While many solutions have been proposed for this task, they are usually not optimized even for common smartphone AI hardware, not to mention more constrained smart TV platforms that are often supporting INT8 inference only. To address this problem, we introduce the first Mobile AI challenge, where the target is to develop an end-to-end deep learning-based image super-resolution solutions that can demonstrate a real-time performance on mobile or edge NPUs. For this, the participants were provided with the DIV2K dataset and trained quantized models to do an efficient 3X image upscaling. The runtime of all models was evaluated on the Synaptics VS680 Smart Home board with a dedicated NPU capable of accelerating quantized neural networks. The proposed solutions are fully compatible with all major mobile AI accelerators and are capable of reconstructing Full HD images under 40-60 ms while achieving high fidelity results. A detailed description of all models developed in the challenge is provided in this paper.

\end{abstract}
{\let\thefootnote\relax\footnotetext{%
\hspace{-5mm}$^*$
Andrey Ignatov, Radu Timofte, Maurizio Denna and Abdel Younes are the Mobile AI 2021 challenge organizers \textit{(andrey@vision.ee.ethz.ch, radu.timofte@vision.ee.ethz.ch, maurizio.denna@synaptics.com, abdel. younes@synaptics.com)}. The other authors participated in the challenge. \\ Appendix \ref{sec:apd:team} contains the authors' team names and affiliations. \vspace{2mm} \\ Mobile AI 2021 Workshop website: \\ \url{https://ai-benchmark.com/workshops/mai/2021/}
}}

\section{Introduction}

\begin{figure*}[t!]
\centering
\setlength{\tabcolsep}{1pt}
\resizebox{\linewidth}{!}
{
\includegraphics[width=0.5\linewidth]{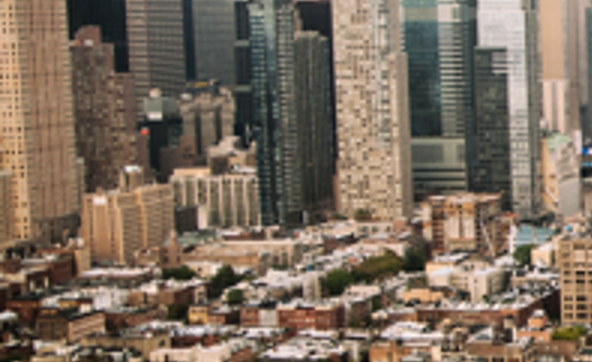} \hspace{2mm}
\includegraphics[width=0.5\linewidth]{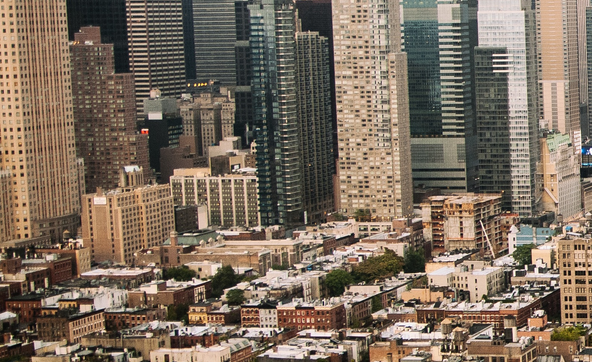}
}
\vspace{0cm}
\caption{Sample crops from a 3X bicubically upscaled image and the target DIV2K~\cite{agustsson2017ntire} photo.}
\label{fig:example_photos}
\end{figure*}

Image super-resolution is a classical computer vision problem where the goal is to reconstruct the original image based on its downscaled version, adding the lost lost high frequencies and rich texture details. During the past years, this task has witnessed an increased popularity due to its direct application to telephoto image processing in smartphone cameras, low-resolution media data enhancement as well as to upscaling images and videos to the target high resolution of display panels. Numerous classical~\cite{irani1991improving,freeman2002example,park2003super,timofte2013anchored,timofte2014a+,yang2013fast,yang2008image,yang2010image,huang2015single,timofte2016seven} and deep learning-based~\cite{dong2015image,dong2014learning,kim2016accurate,lim2017enhanced,timofte2017ntire,timofte2018ntire,cai2019ntire,lugmayr2020ntire,zhang2020ntire,ignatov2018pirm} approaches have been proposed for this task in the past. The biggest limitation of these methods is that they were primarily targeted at achieving high fidelity scores while not optimized for computational efficiency and mobile-related constraints, which is essential for this and other tasks related to image processing and enhancement~\cite{ignatov2017dslr,ignatov2018wespe,ignatov2020replacing} on mobile devices. In this challenge, we take one step further in solving this problem by using a popular DIV2K~\cite{agustsson2017ntire} image super-resolution dataset and by imposing additional efficiency-related constraints on the developed solutions.

When it comes to the deployment of AI-based solutions on mobile devices, one needs to take care of the particularities of mobile NPUs and DSPs to design an efficient model. An extensive overview of smartphone AI acceleration hardware and its performance is provided in~\cite{ignatov2019ai,ignatov2018ai}. According to the results reported in these papers, the latest mobile NPUs are already approaching the results of mid-range desktop GPUs released not long ago. However, there are still two major issues that prevent a straightforward deployment of neural networks on mobile devices: a restricted amount of RAM, and a limited and not always efficient support for many common deep learning layers and operators. These two problems make it impossible to process high resolution data with standard NN models, thus requiring a careful adaptation of each architecture to the restrictions of mobile AI hardware. Such optimizations can include network pruning and compression~\cite{chiang2020deploying,ignatov2020rendering,li2019learning,liu2019metapruning,obukhov2020t}, 16-bit / 8-bit~\cite{chiang2020deploying,jain2019trained,jacob2018quantization,yang2019quantization} and low-bit~\cite{cai2020zeroq,uhlich2019mixed,ignatov2020controlling,liu2018bi} quantization, device- or NPU-specific adaptations, platform-aware neural architecture search~\cite{howard2019searching,tan2019mnasnet,wu2019fbnet,wan2020fbnetv2}, \etc.

While many challenges and works targeted at efficient deep learning models have been proposed recently, the evaluation of the obtained solutions is generally performed on desktop CPUs and GPUs, making the developed solutions not practical due to the above mentioned issues. To address this problem, we introduce the first \textit{Mobile AI Workshop and Challenges}, where all deep learning solutions are developed for and evaluated on real mobile devices.
In this competition, the participating teams were provided with the DIV2K~\cite{agustsson2017ntire} dataset containing diverse 2K resolution RGB images used to train their models using a downscaling factor of 3. More importantly, since many mobile and smart TV platforms can accelerate only INT8 models, all submitted solutions had to be fully-quantized.
Within the challenge, the participants were evaluating the runtime and tuning their models on the Synaptics Dolphin platform featuring a dedicated NPU that can efficiently accelerate INT8 neural networks.
The final score of each submitted solution was based on the runtime and fidelity results, thus balancing between the image reconstruction quality and efficiency of the proposed model. Finally, all developed solutions are fully compatible with the TensorFlow Lite framework~\cite{TensorFlowLite2021}, thus can be deployed and accelerated on any mobile platform providing AI acceleration through the Android Neural Networks API (NNAPI)~\cite{NNAPI2021} or custom TFLite delegates~\cite{TFLiteDelegates2021}.

\smallskip


This challenge is a part of the \textit{MAI 2021 Workshop and Challenges} consisting of the following competitions:


\small

\begin{itemize}
\item Learned Smartphone ISP on Mobile NPUs~\cite{ignatov2021learned}
\item Real Image Denoising on Mobile GPUs~\cite{ignatov2021fastDenoising}
\item Quantized Image Super-Resolution on Edge SoC NPUs
\item Real-Time Video Super-Resolution on Mobile GPUs~\cite{romero2021real}
\item Single-Image Depth Estimation on Mobile Devices~\cite{ignatov2021fastDepth}
\item Quantized Camera Scene Detection on Smartphones~\cite{ignatov2021fastSceneDetection}
\item High Dynamic Range Image Processing on Mobile NPUs
\end{itemize}

\normalsize


\noindent The results obtained in the other competitions and the description of the proposed solutions can be found in the corresponding challenge papers.


\begin{figure*}[t!]
\centering
\setlength{\tabcolsep}{1pt}
\resizebox{0.96\linewidth}{!}
{
\includegraphics[width=1.0\linewidth]{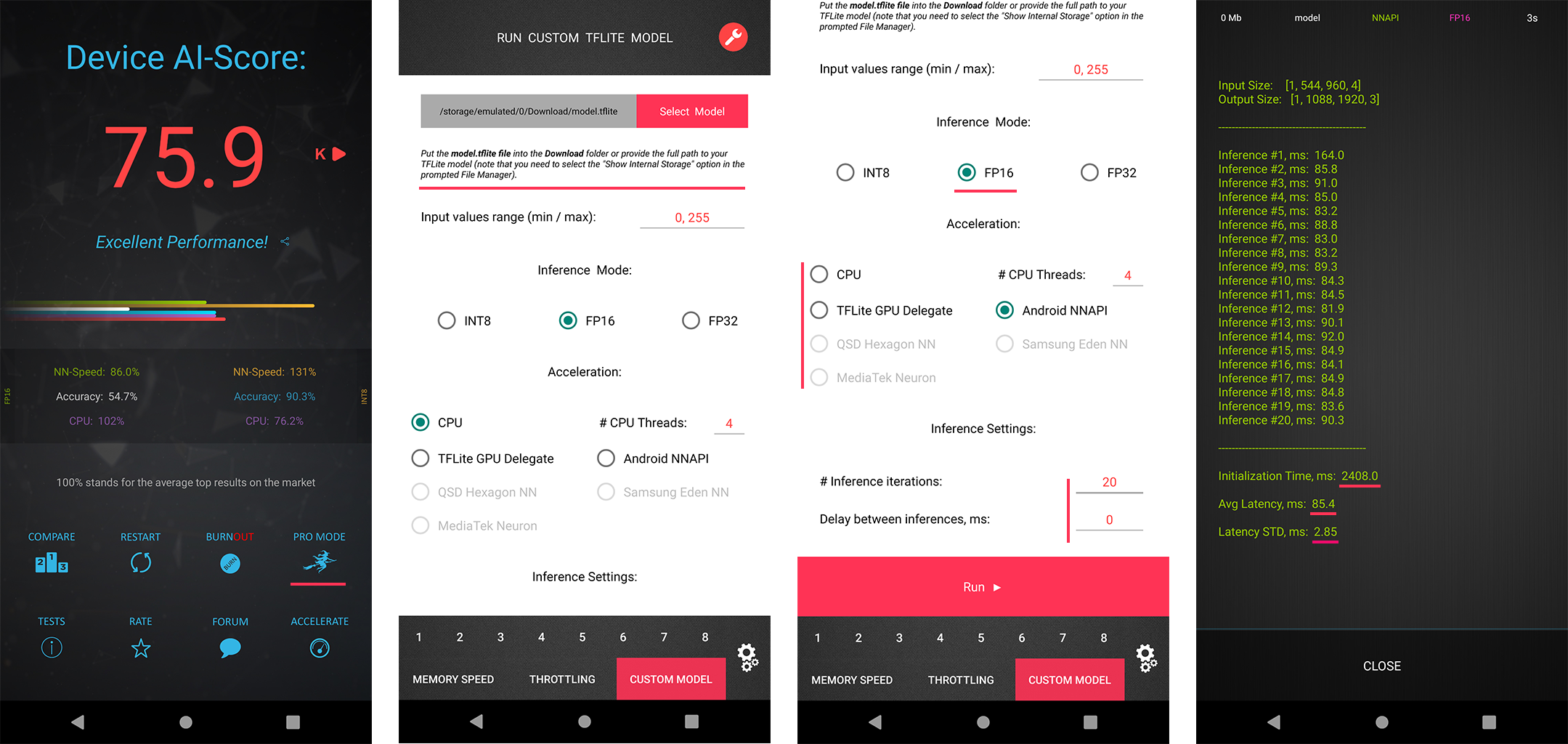}
}
\vspace{0.2cm}
\caption{Loading and running custom TensorFlow Lite models with AI Benchmark application. The currently supported acceleration options include Android NNAPI, TFLite GPU, Hexagon NN, Samsung Eden and MediaTek Neuron delegates as well as CPU inference through TFLite or XNNPACK backends. The latest app version can be downloaded at \url{https://ai-benchmark.com/download}}
\label{fig:ai_benchmark_custom}
\end{figure*}

\section{Challenge}

To develop an efficient and practical solution for mobile-related tasks, one needs the following major components:

\begin{enumerate}
\item A high-quality and large-scale dataset that can be used to train and evaluate the solution;
\item An efficient way to check the runtime and debug the model locally without any constraints;
\item An ability to regularly test the runtime of the designed neural network on the target mobile platform or device.
\end{enumerate}

This challenge addresses all the above issues. Real training data, tools, and runtime evaluation options provided to the challenge participants are described in the next sections.

\subsection{Dataset}

In this challenge, the participants were proposed to work with the popular DIV2K~\cite{agustsson2017ntire} dataset. It consists from 1000 divers 2K resolution RGB images: 800 are used for training, 100 for validation and 100  for testing purposes. The images are of high quality both aesthetically and in the terms of small amounts of noise and other corruptions (like blur and color shifts). All images were manually collected and have 2K pixels on at least one of the axes (vertical or horizontal). DIV2K covers a large diversity of contents, from people, handmade objects and environments (cities), to  flora and fauna and natural sceneries, including underwater. An example set of images is demonstrated in Fig.~\ref{fig:example_photos}.

\subsection{Local Runtime Evaluation}

When developing AI solutions for mobile devices, it is vital to be able to test the designed models and debug all emerging issues locally on available devices. For this, the participants were provided with the \textit{AI Benchmark} application~\cite{ignatov2018ai,ignatov2019ai} that allows to load any custom TensorFlow Lite model and run it on any Android device with all supported acceleration options. This tool contains the latest versions of \textit{Android NNAPI, TFLite GPU, Hexagon NN, Samsung Eden} and \textit{MediaTek Neuron} delegates, therefore supporting all current mobile platforms and providing the users with the ability to execute neural networks on smartphone NPUs, APUs, DSPs, GPUs and CPUs.

\smallskip

To load and run a custom TensorFlow Lite model, one needs to follow the next steps:

\begin{enumerate}
\setlength\itemsep{0mm}
\item Download AI Benchmark from the official website\footnote{\url{https://ai-benchmark.com/download}} or from the Google Play\footnote{\url{https://play.google.com/store/apps/details?id=org.benchmark.demo}} and run its standard tests.
\item After the end of the tests, enter the \textit{PRO Mode} and select the \textit{Custom Model} tab there.
\item Rename the exported TFLite model to \textit{model.tflite} and put it into the \textit{Download} folder of the device.
\item Select mode type \textit{(INT8, FP16, or FP32)}, the desired acceleration/inference options and run the model.
\end{enumerate}

\noindent These steps are also illustrated in Fig.~\ref{fig:ai_benchmark_custom}.

\subsection{Runtime Evaluation on the Target Platform}

In this challenge, we use the \textit{Synaptics VS680 Edge AI SoC} \cite{SynapticsSoC2021} Evaluation Kit as our target runtime evaluation platform. The VS680 Edge AI SoC is integrated into Smart Home solution and it features a powerful NPU designed by \textit{VeriSilicon} and capable of accelerating quantized models (up to 7 TOPS). It supports Android and can perform NN inference through NNAPI, demonstrating INT8 AI Benchmark scores that are close to the ones of mid-range smartphone chipsets. Within the challenge, the participants were able to upload their TFLite models to an external server and get feedback regarding the speed of their model: the inference time of their solution on the above mentioned NPU or an error log if the network contained incompatible operations and/or improper quantization. Participants' models were parsed and accelerated using Synaptics' TFLite delegate that can dynamically parse and map a given model's higher level representation of neural network layers and operations to its equivalent internal binary representation that will be optimized for efficient integer only execution on the VS680's NPU. The same setup was also used for the final runtime evaluation. The participants were additionally provided with a list of ops supported by this board and model optimization guidance in order to fully utilize the NPU's convolution and tensor processing resources.

\begin{table*}[t!]
\centering
\resizebox{\linewidth}{!}
{
\begin{tabular}{l|c|cc|cc|cc|ccc|c}
\hline
Team \, & \, Author \, & \, Framework \, & Model Size, & \, PSNR$\uparrow$ \, & \, SSIM$\uparrow$ \, & \, $\Delta$ PSNR \, & \, $\Delta$ SSIM \, & \multicolumn{2}{c}{\, Runtime, ms $\downarrow$ \,} & Speed-Up & \, Final Score \\
& & & \small{KB} & \multicolumn{2}{c|}{\small{INT8 Model}} & \multicolumn{2}{c|}{\small{FP32 $\to$ INT8 Acc. Drop}} & \, \small{CPU} \, & \, \small{NPU} \, & & \\
\hline
\hline
Aselsan Research & deepernewbie & Keras / TensorFlow & 67 & 29.58 & 0.86 & 0.18 & 0.0093 & 1278 & 44.85 & 28.5 & \textBF{51.02} \\
Noah\_TerminalVision & JeremieG & Keras / TensorFlow & 109 & 29.41 & 0.8537 & 0.33 & 0.0142 & 668 & \textBF{38.32} & 17.4 & 47.18 \\
ALONG & richlaji & TensorFlow & 30 & 29.52 & 0.8607 & \footnotesize{N.A.} & \footnotesize{N.A.} & 951 & 62.25 & 15.3 & 33.82 \\
\hline
\rowcolor{grayhighlight} A+ regression~\cite{timofte2014a+} & Baseline &  &  & 29.32 & 0.8520 & - & - & - & - & - & - \\
\hline
EmbededAI & xindongzhang & PyTorch / TensorFlow & 82 & 28.82 & 0.8428 & 0.15 & 0.0119 & 1224 & 76.61 & 16 & 10.41 \\
mju\_gogogo & mju\_gogogo & Keras / TensorFlow & 940 & 28.92 & 0.8486 & \footnotesize{N.A.} & \footnotesize{N.A.} & \footnotesize{\textcolor{red}{Failed}} & 718 & - & 1.28 \\
\hline
\rowcolor{grayhighlight} Bicubic Upscaling & Baseline &  &  & 28.26 & 0.8277 & - & - & - & - & - & - \\
\hline
221B & masanshu & TensorFlow & 175 & 25.44 & 0.729 & 4.03 & 0.1337 & \footnotesize{\textcolor{red}{Failed}} & 238.43 & - & 0.03 \\
svnit\_ntnu & kalpesh\_svnit & TensorFlow & 8 & 19.3 & 0.7061 & 9.61 & 0.1442 & 1947 & 78.84 & 24.7 & 0.00 \\
CVML & vishalchudasama & TensorFlow & 10 & 19.5 & 0.7462 & 9.45 & 0.1049 & 1772 & 90.20 & 19.6 & 0.00 \\
TieGuoDun Team & ShuaiqiXiaopangzhi & TensorFlow & 636 & 16.19 & 0.6654 & 13.28 & 0.1991 & \footnotesize{\textcolor{red}{Failed}} & 913.96 & - & 0.00 \\
\hline
\textit{MCG} $^*$ & \textit{TinyJie} & TensorFlow & 53 & \textBF{29.87} & \textBF{0.8686} & \footnotesize{N.A.} & \footnotesize{N.A.} & 998 & 36.89 & 27 & \textBF{92.72} \\
\end{tabular}
}
\vspace{2.6mm}
\caption{\small{Mobile AI 2021 Real-Time Image Super-Resolution challenge results and final rankings. During the runtime measurements, the models were performing image upscaling from 640$\times$360 to 1920$\times$1080 pixels. $\Delta$ PSNR and $\Delta$ SSIM values correspond to accuracy loss measured in comparison to the original floating-point network. Team \textit{Aselsan Research} is the challenge winner. $^*$~The presented solution from \textit{TinyJie} team was submitted after the official challenge deadline.}}
\label{tab:results}
\end{table*}

\subsection{Challenge Phases}

The challenge consisted of the following phases:

\vspace{-0.8mm}
\begin{enumerate}
\item[I.] \textit{Development:} the participants get access to the data and AI Benchmark app, and are able to train the models and evaluate their runtime locally;
\item[II.] \textit{Validation:} the participants can upload their models to the remote server to check the fidelity scores on the validation dataset, to get the runtime on the target platform, and to compare their results on the validation leaderboard;
\item[III.] \textit{Testing:} the participants submit their final results, codes, TensorFlow Lite models, and factsheets.
\end{enumerate}
\vspace{-0.8mm}

\subsection{Scoring System}

All solutions were evaluated using the following metrics:

\vspace{-0.8mm}
\begin{itemize}
\setlength\itemsep{-0.2mm}
\item Peak Signal-to-Noise Ratio (PSNR) measuring fidelity score,
\item Structural Similarity Index Measure (SSIM), a proxy for perceptual score,
\item The runtime on the target Synaptics VS680 board.
\end{itemize}
\vspace{-0.8mm}

The score of each final submission was evaluated based on the next formula ($C$ is a constant normalization factor):

\smallskip
\begin{equation*}
\text{Final Score} \,=\, \frac{2^{2 \cdot \text{PSNR}}}{C \cdot \text{runtime}},
\end{equation*}
\smallskip

During the final challenge phase, the participants did not have access to the test dataset. Instead, they had to submit their final TensorFlow Lite models that were subsequently used by the challenge organizers to check both the runtime and the fidelity results of each submission under identical conditions. This approach solved all the issues related to model overfitting, reproducibility of the results, and consistency of the obtained runtime/accuracy values.

\section{Challenge Results}

From above 180 registered participants, 12 teams entered the final phase and submitted their results, TFLite models, codes, executables and factsheets. Table~\ref{tab:results} summarizes the final challenge results and reports PSNR, SSIM and runtime numbers for the valid solutions on the final test dataset and on the target evaluation platform. The proposed methods are described in section~\ref{sec:solutions}, and the team members and affiliations are listed in Appendix~\ref{sec:apd:team}.

\begin{figure*}[t!]
\centering
\resizebox{0.9\linewidth}{!}
{
\includegraphics[width=1.0\linewidth]{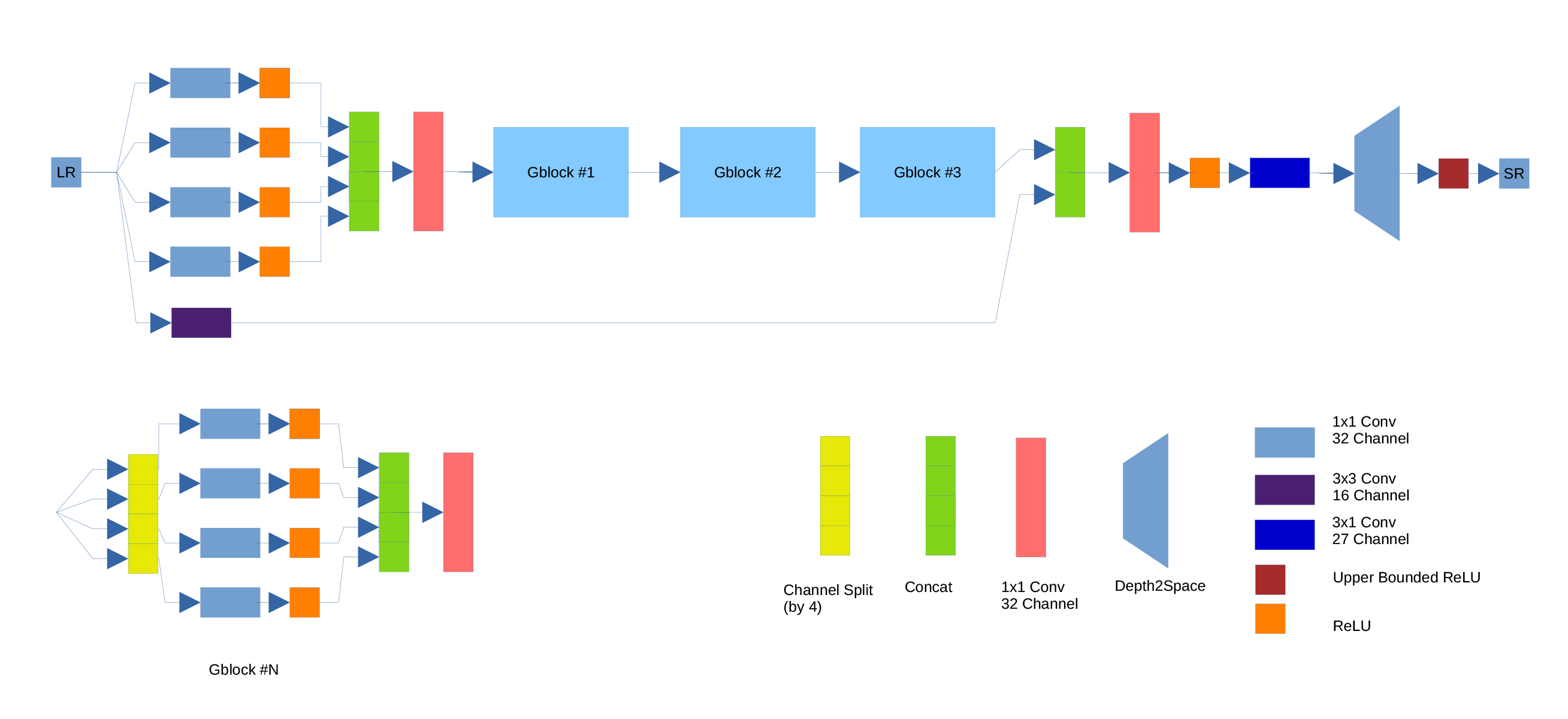}
}
\caption{\small{The model architecture proposed by Aselsan Research team.}}
\label{fig:Aselsan}
\end{figure*}

\subsection{Results and Discussion}

The problem considered in this competition was very challenging as the solutions had to be both optimized for the target Smart Home platform and be fully quantized. While there exists many works proposing various weight quantization techniques, the majority of them are still using floating-point activations: in this case, the resulting models are not compatible with INT8 NPUs, and the entire idea of network quantization is lost as no benefits are obtained on real hardware. Therefore, in this challenge the participants had to perform full model quantization including inputs, weights, convolutions and activations~-- networks with floating-point ops were not accepted. As one can see (Fig.~\ref{tab:results}), only 6 teams were able to outperform a simple bicubic image upsampling baseline, in all other cases the accuracy loss after quantization was enormous, and the models were just producing corrupted outputs. One major issue met by the majority of challenge participants was to avoid using floating-point output dequantization: without this block, quantized outputs were often normalized incorrectly, resulting in wrong network's output values scaling. The easiest way to deal with this problem was to add a simple clipped \textit{ReLU} layer on top of the model, then the outputs were mapped linearly to the [0, 255] interval. Applying quantized-aware training was also helping a lot in getting good fidelity results.

The majority of the proposed solutions demonstrated a very high efficiency, being able to upscale 640$\times$360 pixel input images to Full HD resolution under 60-80 ms on the target Synaptics VS680 board. Since not all TFLite operations were equally optimized by the target platform, participants had to rely on a recommended set of ops for building their models in order to design a solution that would maximize NPU utilization. Team \textit{Aselsan Research} is the challenge winner~--- the authors were able to achieve good fidelity and runtime values by using a relatively small model with grouped convolutions. Even better results were obtained by team \textit{MCG}, though, unfortunately, it was able to solve all issues related to model quantization only after the end of the challenge. This team as well as \textit{Noah\_TerminalVision} and \textit{ALONG} were using a similar idea of learning only SR residuals with convolutional or residual blocks. Quantized models submitted by \textit{EmbededAI} and \textit{mju\_gogogo} demonstrated only a slight improvement over the baseline bicubic upscaling approach. The rest of the teams were not able to fight the accuracy drop resulted from model quantization, though their original floating-point networks were showing quite good fidelity scores.

At the beginning of the runtime validation phase, almost all submitted solutions were either crushing on the target Synaptics platform or demonstrating a runtime of several seconds when tested on the target resolution images. It took a large number of iterations for the majority of teams to come up with solutions that can be accelerated efficiently on the considered NPU, though many models were very light and demonstrated good speed on desktop CPUs and GPUs from the very beginning. This explicitly shows that the runtime values obtained on common deep learning hardware are not representative when it comes to model deployment on mobile AI silicon: even solutions that might seem to be very efficient can struggle significantly due to the specific constraints of IoT and mobile AI acceleration hardware and frameworks. This makes deep learning development for mobile devices so challenging, though the results obtained in this competition demonstrate that one can get a very efficient model when taking the above aspects into account.

\section{Challenge Methods}
\label{sec:solutions}

\noindent This section describes solutions submitted by all teams participating in the final stage of the MAI 2021 Real-Time Image Super-Resolution challenge.

\subsection{Aselsan Research}

The architecture proposed by Aselsan Research is presented in Fig.~\ref{fig:Aselsan}. The major building block (Gblock) of this model is based on the concept of grouped convolutions: the input feature maps are split into 4 parts and fed to separate convolutions (working in parallel) to decrease the RAM consumption and computational costs. The authors emphasize that though replacing the standard convolutional layers with separable convolutions results in better runtime, this also leads to a large accuracy drop after performing model quantization, thus they were not used in the final architecture. An additional skip connection was added to improve the fidelity results of the proposed solution, no input data normalization was used to increase the speed of the model.

The network was trained on 32$\times$32 pixel input images with a batch size of 16. \textit{Charbonnier} loss function was minimized using Adam optimizer with a dynamic learning rate ranging from $25e-4$ to $1e-4$. Model quantization was performed with TensorFlow's standard post-training quantization utilities, clipped \textit{ReLU} was added on top of the model to avoid incorrect output normalization. A more detailed description of the model, design choices and training procedure is provided in~\cite{ayazoglu2021extremely}.

\subsection{MCG}

\begin{figure}[h!]
\centering
\resizebox{1.0\linewidth}{!}
{
\includegraphics[width=1.0\linewidth]{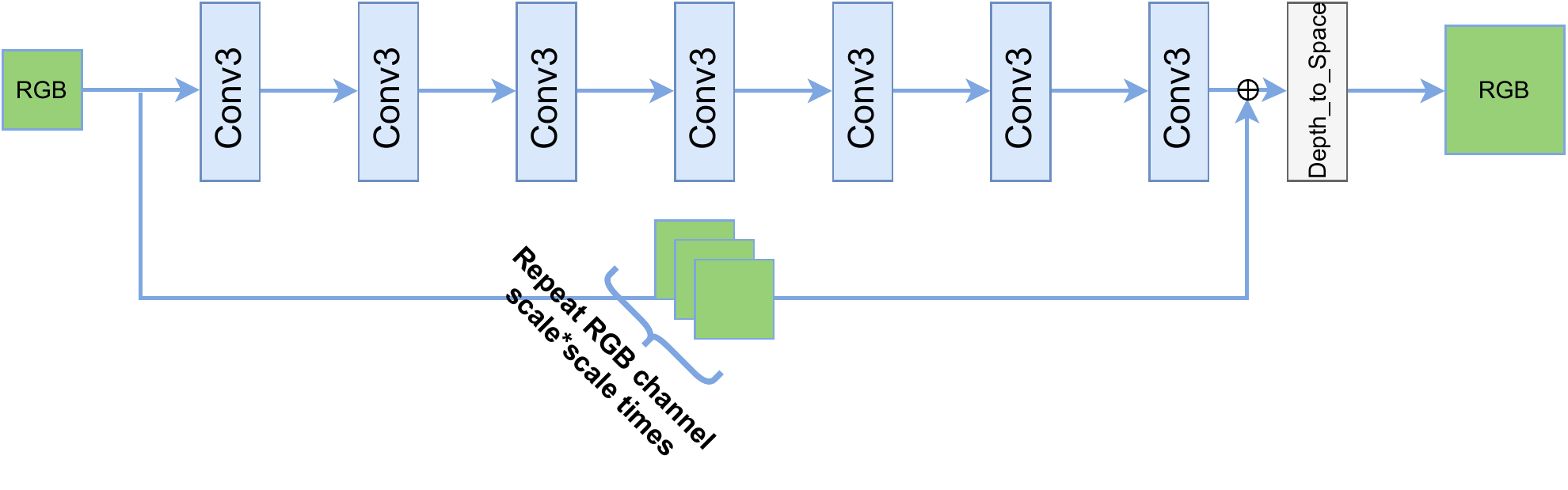}
}
\caption{\small{Anchor-based CNN proposed by MCG team.}}
\label{fig:MCG}
\end{figure}

Team MCG proposed an anchor-based CNN (Fig.~\ref{fig:MCG}) for the considered problem. The main idea behind this architecture is to learn only the residual part of SR image. If we remove all convolutional layers from this model, then the workflow would be as follows: the input image is stacked 3$\times$3 = 9 times (where 3 is the upscaling factor) and then reshaped by the \textit{depth-to-space} layer to the target resolution. The resulting image will have the same size as the target SR one~--- resizing is achieved by repeating each pixel value 9 times. The ``removed'' convolutional block is therefore learning the difference between the low-resolution and SR photos, which is added to the input image before the \textit{depth-to-space} layer. This block consists of five 3$\times$3 convolutional layers followed by \textit{ReLU} activations, and one additional conv layer on top of them.

The network was trained with a batch size of 16 on 64$\times$64 pixel input images augmented by random flipping and rotation. $L_1$ loss was used as a target metric, model parameters were optimized for 1000 epochs using Adam with a learning rate initialized at $1e-3$ and decreases by half every 200 epochs. Quantized-aware training as well as post-training quantization were applied to get an accurate INT8 model, clipped \textit{ReLU} was added on top of the network to avoid incorrect output normalization. A detailed description of the proposed method is also provided in~\cite{du2021anchor}.

\subsection{Noah\_TerminalVision}

\begin{figure}[h!]
\centering
\resizebox{1.0\linewidth}{!}
{
\includegraphics[width=1.0\linewidth]{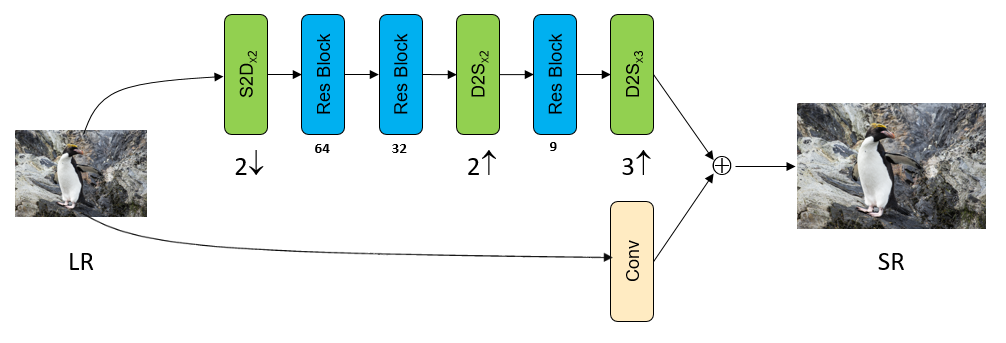}
}
\caption{\small{TinySRNet model designed by Noah\_TerminalVision.}}
\label{fig:Noah}
\end{figure}

Team Noah\_TerminalVision developed a small TinySRNet model demonstrated in Fig.~\ref{fig:Noah}. This network contains three residual blocks (each consisting of two convolutions), space-to-depth (S2D), depth-to-space (D2S) and one residual convolutional layer. The authors especially emphasize the importance of the residual block which helps a lot in maintaining good accuracy after model quantization. The network was trained to minimize $L_1$ loss, its parameters were optimized using Adam for one million iterations with a cyclic learning rate starting from $5e-4$ and decreased to $1e-6$ each 200K iterations. Quantized-aware training was applied to improve the accuracy of the resulting INT8 model.

\subsection{ALONG}

\begin{figure}[h!]
\centering
\resizebox{1.0\linewidth}{!}
{
\includegraphics[width=1.0\linewidth]{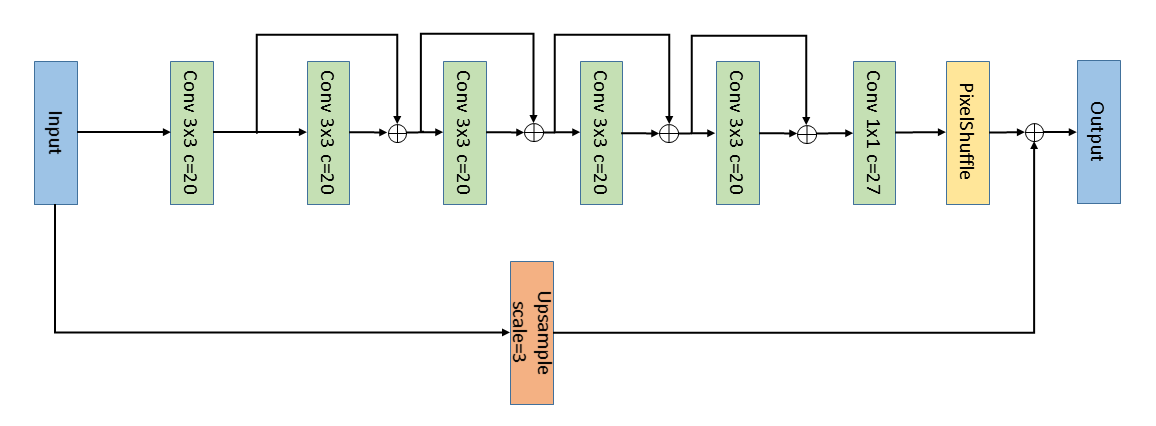}
}
\caption{\small{The model architecture proposed by ALONG team.}}
\label{fig:ALONG}
\end{figure}

The architecture developed by team ALONG (Fig.~\ref{fig:ALONG}) is very similar to the previous solution, the major difference consists in doing all processing on the original scale and using nearest neighbor upsampling instead of convolution in the residual block connecting the input and output layers. The model was first trained to minimize $L_1$ loss function and then fine-tuned with $L_2$ loss. 128$\times$128px input patches augmented with
random flips and rotations were used during the training, model parameters were optimized using Adam with an initial learning rate of $2e-4$ halved every 200K iterations. Quantized-aware training was applied to improve the accuracy of the resulting INT8 model.

\subsection{EmbededAI}

\begin{figure}[h!]
\centering
\resizebox{0.84\linewidth}{!}
{
\includegraphics[width=1.0\linewidth]{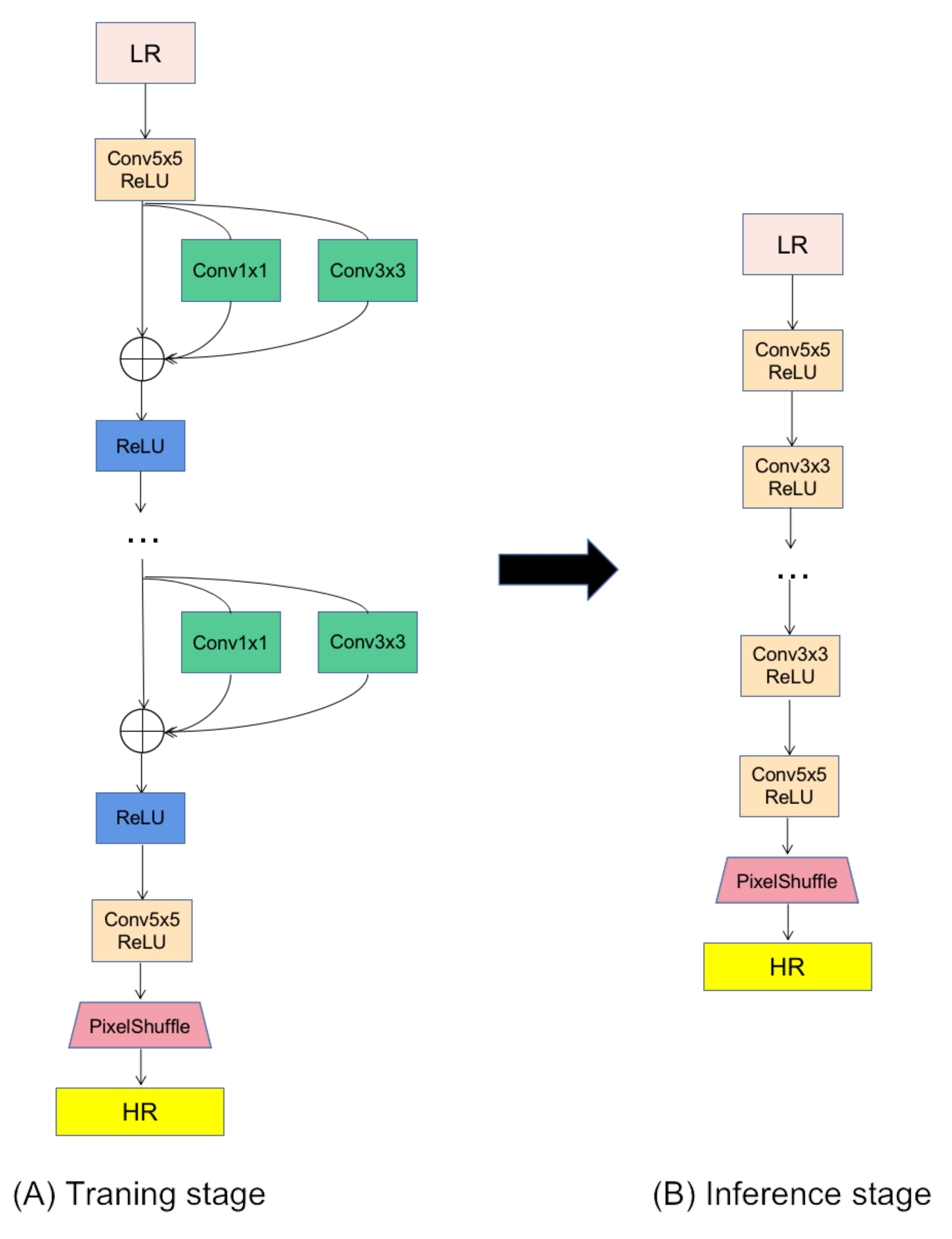}
}
\caption{\small{Network topology of the PRPSR CNN developed by EmbededAI team during training and inference stages.}}
\label{fig:EmbededAI}
\end{figure}

Team EmbededAI presented a Plain Re-Parameterizable Convolutions for Super Resolution (PRPSR) model for the considered task. This network is designed using a pure convolutional topology: the input low-resolution image is fed to 5$\times$5 convolutional layer performing feature extraction, followed by five 3$\times$3 convolutions, one 3$\times$3 convolutional and one pixel-shuffle layer reshaping the output to the target resolution (Fig.~\ref{fig:EmbededAI}, right image). No residual blocks, skip connections, up-sampling or even addition operations are used in the model to improve its speed. During the training stage, each single convolutional layer is decoupled into a group of three operations:
\begin{equation*}
y = x + conv_{1\times1}(x) + conv_{3\times3}(x),
\end{equation*}
followed by \textit{ReLU} activations (Fig.~\ref{fig:EmbededAI}, left image). After the end of the training, the parameters of each convolutional group are folded to get a single convolution~\cite{ding2021repvgg}.

The network was trained with a batch size of 32 on 64$\times$64 pixel input images. $L_1$ loss was used as a target metric, model parameters were optimized for 600 epochs using Adam with an initial learning rate of $5e-4$ decreases by half every 200 epochs. Model quantization was performed with TensorFlow's post-training quantization tools, clipped \textit{ReLU} was used after the last convolutional layer to avoid incorrect outputs normalization.

\subsection{mju\_gogogo}

\begin{figure}[h!]
\centering
\resizebox{1.0\linewidth}{!}
{
\includegraphics[width=1.0\linewidth]{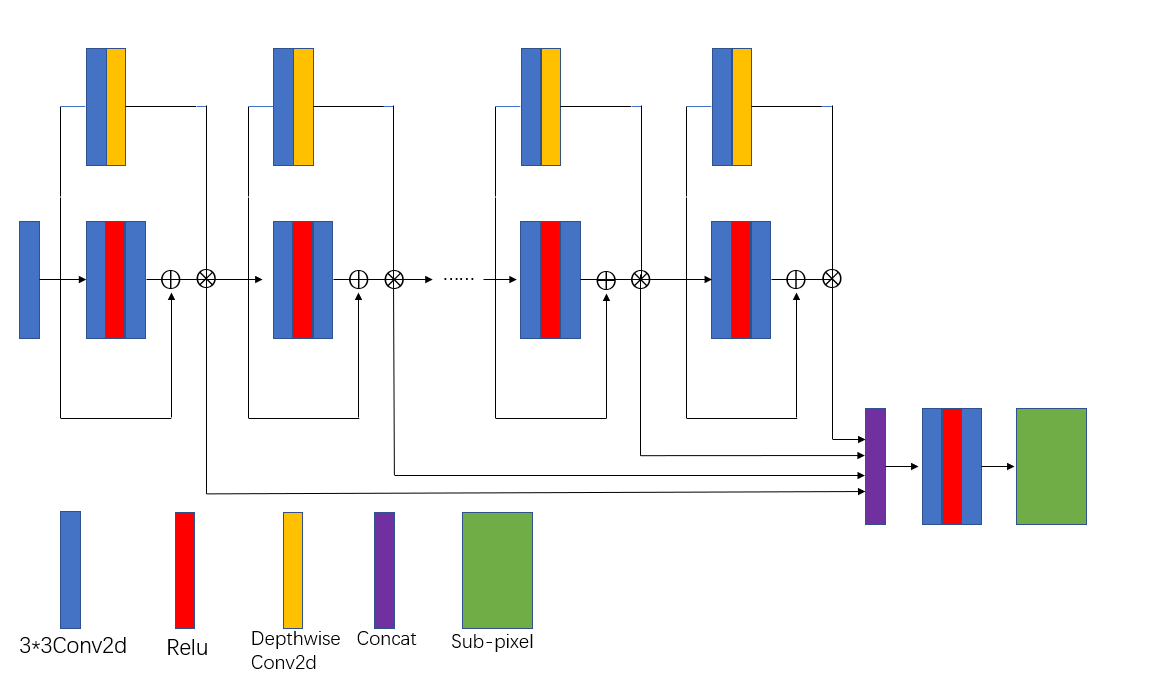}
}
\caption{\small{EDSR-based model with attention blocks proposed by mju\_gogogo team.}}
\label{fig:mjugogogo}
\end{figure}

The model architecture proposed by mju\_gogogo team is presented in Fig.~\ref{fig:mjugogogo}. The model is based on the EDSR~\cite{lim2017enhanced} design with additional spatial attention blocks. The network is relatively large~--- it consists of six residual blocks, each one multiplied by the outputs of the corresponding attention unit, one 1$\times$1 convolutional layer reducing the number of filters from 384 to 64, one global skip connection, three 3$\times$3 convolutions and one \textit{depth-to-space} layer. The model was first trained to minimize the \textit{mean absolute error (MAE)} with Adam optimizer for 750 epochs, and then fine-tuned using quantized-aware training and the \textit{MSE} loss for another 75 epochs.

\section{Additional Literature}

An overview of the past challenges on mobile-related tasks together with the proposed solutions can be found in the following papers:

\begin{itemize}
\item Image Super-Resolution:\, \cite{ignatov2018pirm,lugmayr2020ntire,cai2019ntire,timofte2018ntire}
\item Learned End-to-End ISP:\, \cite{ignatov2019aim,ignatov2020aim}
\item Perceptual Image Enhancement:\, \cite{ignatov2018pirm,ignatov2019ntire}
\item Bokeh Effect Rendering:\, \cite{ignatov2019aimBokeh,ignatov2020aimBokeh}
\item Image Denoising:\, \cite{abdelhamed2020ntire,abdelhamed2019ntire}
\end{itemize}

\section*{Acknowledgements}

We thank Synaptics Inc., AI Witchlabs and ETH Zurich (Computer Vision Lab), the organizers and sponsors of this Mobile AI 2021 challenge.

\appendix
\section{Teams and Affiliations}
\label{sec:apd:team}

\bigskip

\subsection*{Mobile AI 2021 Team}
\noindent\textit{\textbf{Title: }}\\ Mobile AI 2021 Image Super-Resolution Challenge\\
\noindent\textit{\textbf{Members:}}\\ Andrey Ignatov$^{1,4}$ \textit{(andrey@vision.ee.ethz.ch)}, Radu Timofte$^{1,4}$  \textit{(radu.timofte@vision.ee.ethz.ch)}, Maurizio Denna$^2$ \textit{(maurizio.denna@synaptics.com)}, Abdel Younes$^2$ \textit{(abdel.younes@synaptics.com)}, Andrew Lek$^3$ \textit{(andrew.lek@synaptics.com)}\\
\noindent\textit{\textbf{Affiliations: }}\\
$^1$ Computer Vision Lab, ETH Zurich, Switzerland\\
$^2$ Synaptics Europe, Switzerland\\
$^3$ Synaptics HQ, San Jose, California, USA\\
$^4$ AI Witchlabs, Switzerland\\

\subsection*{Aselsan Research}
\noindent\textit{\textbf{Title:}}\\Extremely Lightweight Quantization Robust Real-Time Single-Image Super Resolution for Mobile Devices~\cite{ayazoglu2021extremely}\\
\noindent\textit{\textbf{Members:}}\\ \textit{Mustafa Ayazoglu (mayazoglu@aselsan.com.tr)}\\
\noindent\textit{\textbf{Affiliations: }}\\
Aselsan Corporation, Turkey\\
\url{https://www.aselsan.com.tr/}\\

\subsection*{MCG}
\noindent\textit{\textbf{Title:}}\\Anchor-based Net for Mobile Image Super-Resolution~\cite{du2021anchor}\\
\noindent\textit{\textbf{Members:}}\\ \textit{Jie Liu (jieliu@smail.nju.edu.cn)}, Zongcai Du\\
\noindent\textit{\textbf{Affiliations: }}\\
Nanjing University, China\\

\subsection*{Noah\_TerminalVision}
\noindent\textit{\textbf{Title:}}\\TinySRNet for Real-Time Image Super-Resolution\\
\noindent\textit{\textbf{Members:}}\\ \textit{Jiaming Guo (guojiaming5@huawei.com)}, Xueyi Zhou, Hao Jia, Youliang Yan\\
\noindent\textit{\textbf{Affiliations: }}\\
Huawei Technologies Co., Ltd, China\\

\subsection*{ALONG}
\noindent\textit{\textbf{Title:}}\\FastSR: Solution for Real-Time Image Super-Resolution Challenge\\
\noindent\textit{\textbf{Members:}}\\ \textit{Zexin Zhang (m15622188336@163.com)}, Yixin Chen, Yunbo Peng, Yue Lin\\
\noindent\textit{\textbf{Affiliations: }}\\
Netease Games AI Lab, China\\

\subsection*{EmbededAI}
\noindent\textit{\textbf{Title:}}\\Plain Re-parameterizable Convolutions for Efficient Super Resolution\\
\noindent\textit{\textbf{Members:}}\\ \textit{Xindong Zhang (xindongzhang@foxmail.com)}, Hui Zeng\\
\noindent\textit{\textbf{Affiliations: }}\\
The Hong Kong Polytechnic University, Hong Kong\\

\subsection*{mju\_gogogo}
\noindent\textit{\textbf{Title:}}\\EDSR-based Model with Attention Blocks for Image Super-Resolution\\
\noindent\textit{\textbf{Members:}}\\ \textit{Kun Zeng (zengkun301@aliyun.com)}, Peirong Li, Zhihuang Liu, Shiqi Xue, Shengpeng Wang\\
\noindent\textit{\textbf{Affiliations: }}\\
Minjiang University, China\\

{\small
\bibliographystyle{ieee_fullname}

}

\end{document}